\documentclass[a4paper,11pt]{article}
\usepackage[utf8]{inputenc}
\usepackage{graphicx}
\usepackage{amssymb,amsmath,amsthm}
\usepackage{subfigure}
\usepackage[margin=2cm]{geometry}
\usepackage{tikz}
\usepackage{authblk}
\usetikzlibrary{calc,patterns,angles,quotes} 

\usepackage{xcolor}

\title{On interplay between excitability and geometry}
\author{Andrew Adamatzky}
\affil{Unconventional Computing Laboratory,\\ University of the West of England, Bristol BS16 1QY, UK}
\date{}

\graphicspath{{figs/}}

\begin{document}

\maketitle
\begin{abstract}
\noindent
A commonly accepted feature of an excitable medium is that a local excitation leads to a propagation of circular or spiral excitation wave-fronts. This is indeed the case in fully excitable medium. However, with a decrease of an excitability localised wave-fragments emerge and propagate ballistically. Using FitzhHugh-Nagumo  model we numerically study how excitation wave-fronts behave in a geometrically constrained medium and how the wave-fronts explore a random planar graph. We uncover how excitability controls propagation of excitation in angled branches, influences arrest of excitation entering a sudden expansion, and determines patterns of traversing of a random planar graph by an excitation waves. 

\vspace{2mm}

\end{abstract}

\section{Introduction}

Excitation is essential property of all living creatures, from bacteria~\cite{masi2015electrical},  protists~\cite{eckert1979ionic,hansma1979sodium,bingley1966membrane}, fungi~\cite{mcgillviray1987transhyphal} and plants~\cite{trebacz2006electrical,fromm2007electrical,zimmermann2013electrical} to vertebrates~\cite{hodgkin1952quantitative,aidley1998physiology,nelson2012excitable,davidenko1992stationary}. Waves of excitation could be also found in various physical~\cite{kittel1958excitation,tsoi1998excitation,slonczewski1999excitation,gorbunov1987excitation}, chemical~\cite{belousov1959periodic, zhabotinsky1964periodic,zhabotinsky2007belousov} and 
social systems~\cite{farkas2002social,farkas2003human}. 
A study of a propagation of excitation wave-fronts in geometrically constrained media also brings a high value in future and emergent computing technologies because there is a substantial number of theoretical and experimental laboratory prototypes of unconventional computing devices based on the interaction of the  wave-fronts~\cite{adamatzky2005reaction,adamatzky2018reaction}.

Interactions of wave-fronts with inhomogeneities of a medium and geometrical constrains of the medium has been studied for the case of a fully excitable medium, especially in the context of myocardiuam and formation of spiral waves due to inhomogeneities~\cite{goldstein1974changes,kucera1998slow,berenfeld1999dynamics,bub2002propagation,sinha2002critical,kuklik2005reentry,kuklik2010spiral,kogan2010excitation,weise2012emergence,zykov2018wave,wang2018mechanistic,gao2018initiation}. Particularly interesting results include cancellation of  excitation wave-fronts in the narrowing areas~\cite{kogan2010excitation}, annihilation or delay of  wave-fronts propagating along narrow channel when entering a sudden expansion~\cite{kucera1998slow,bub2002propagation}. We reconsider these topics and introduce several new ones in present paper.

A key feature of present study is that we consider behaviour of the medium at a wide range of excitability: from classical fully excitable medium with target waves to sub-excitable medium with localised wave-fragments to non-excitable medium. In numerical experiments we demonstrate dependence of an angle of a branching channel which a wave can enter on the excitability of the medium. We show how the excitability affects arresting of the wave on entering an expansion. We also analyse an extend of a random planar graph exploration by wave-fronts for various values of excitability. 

\section{The model}

FitzHugh-Nagumo (FHN) equations~\cite{fitzhugh1961impulses,nagumo1962active,pertsov1993spiral} give us a qualitative approximation of Hodgkin-Huxley model~\cite{beeler1977reconstruction} of electrical activity of living cells:
\begin{eqnarray}
\frac{\partial v}{\partial t} & = & c_1 u (u-a) (1-u) - c_2 u v + I + D_u \nabla^2 \\
\frac{\partial v}{\partial t} & = & b (u - v)
\end{eqnarray}
where $u$ is a value of a trans-membrane potential, $v$ is a variable accountable for a total slow ionic current, or a recovery variable responsible for a slow negative feedback, $I$ is a value of external stimulation current. Current through intra-cellular spaces is approximated by
$D_u \nabla^2$, where $D_u$ is a conductance. Detailed explanations of the `mechanics' of the model are provided in~\cite{rogers1994collocation}, here we repeat some insights. The term $D_u \nabla^2 u$ governs a passive spread of the current. The terms $c_2 u (u-a) (1-u)$ and $b (u - v)$ describe the ionic currents. The term $u (u-a) (1-u)$ has two stable fixed points $u=0$ and $u=1$ and one unstable point $u=a$, where $a$ is a threshold of an excitation.
We integrated the system using Euler method with five-node Laplace operator, time step $\Delta t=0.015$ and grid point spacing $\Delta x = 2$, other parameters were $D_u=1$, $a=0.13$, $b=0.013$, $c_1=0.26$, we controlled excitability of the medium by varying $c_2$ from 0.09 (fully excitable) to 0.013 (non excitable). Boundaries are considered to be impermeable: $\partial u/\partial \mathbf{n}=0$, where $\mathbf{n}$ is a vector normal to the boundary. 
The numerical integration code, written in Processing  was inspired by \cite{hammer2009, pertsov1993spiral,rogers1994collocation}.
Time-lapse snapshots provided in the paper were recorded at every 150\textsuperscript{th} time step, and we display sites with $u >0.04$; videos supplementing figures were produced by saving a frame of the simulation every 100\textsuperscript{th} step of the numerical integration and assembling them in the video with play rate 30 fps. Videos are available at~\cite{adamatzkyFTHZenodo}.

\section{Results} 

\graphicspath{{figs/FreeSpace/}}
\begin{figure}[!tbp]
    \centering
\subfigure[$c_2=0.1$]{\includegraphics[scale=0.28]{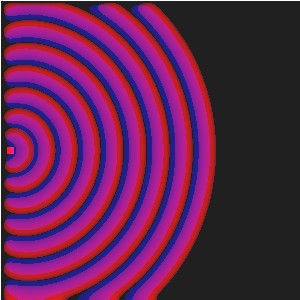}\label{expansion0100000}}
\subfigure[$c_2=0.11$]{\includegraphics[scale=0.28]{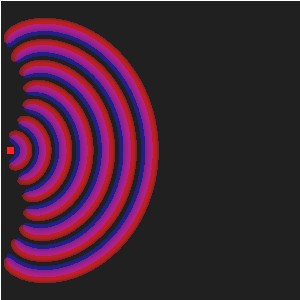}\label{expansion0110000}}
\subfigure[$c_2=0.115$]{\includegraphics[scale=0.28]{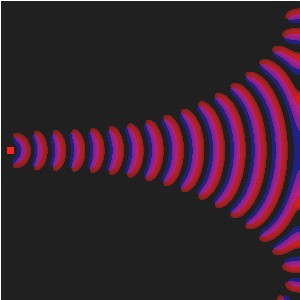}\label{expansion0115000}}
\subfigure[$c_2=0.11505$]{\includegraphics[scale=0.28]{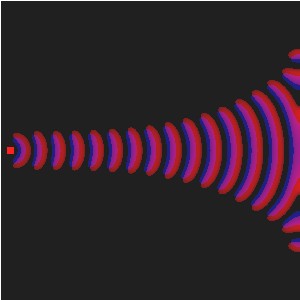}\label{expansion0115050}}
\subfigure[$c_2=0.11508$]{\includegraphics[scale=0.28]{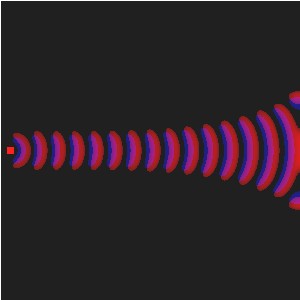}\label{expansion0115080}}
\subfigure[$c_2=0.11509$]{\includegraphics[scale=0.28]{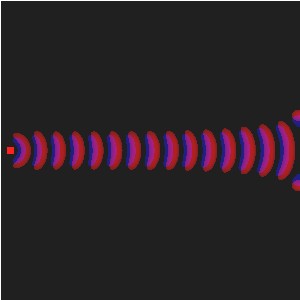}\label{expansion0115090}}
\subfigure[$c_2=0.115092$]{\includegraphics[scale=0.28]{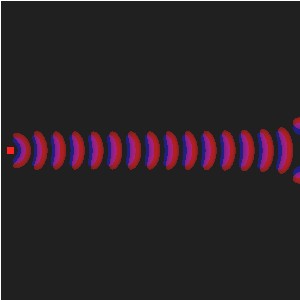}\label{nearsoliton0115092}}
\subfigure[$c_2=0.115093$]{\includegraphics[scale=0.28]{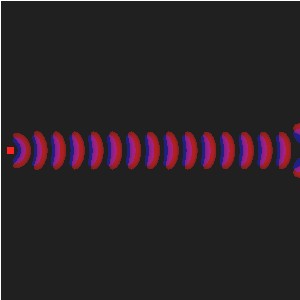}\label{soliton0115093}}
\subfigure[$c_2=0.115095$]{\includegraphics[scale=0.28]{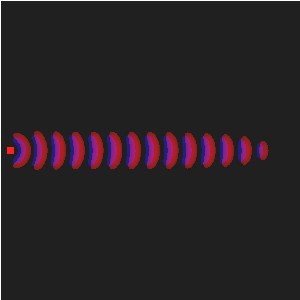}\label{collapse0115095}}
\subfigure[$c_2=0.11510$]{\includegraphics[scale=0.28]{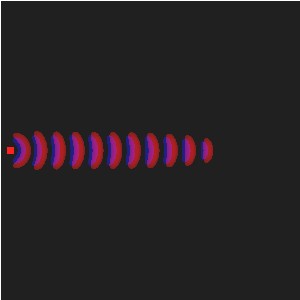}\label{collapse0115100}}
\graphicspath{{figs/}}
\subfigure[$c_2=0.126$]{\includegraphics[scale=0.27]{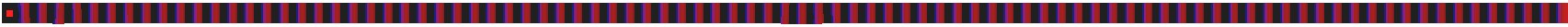}\label{single_channel_lapse_0126}}
\subfigure[]{\includegraphics[scale=0.4]{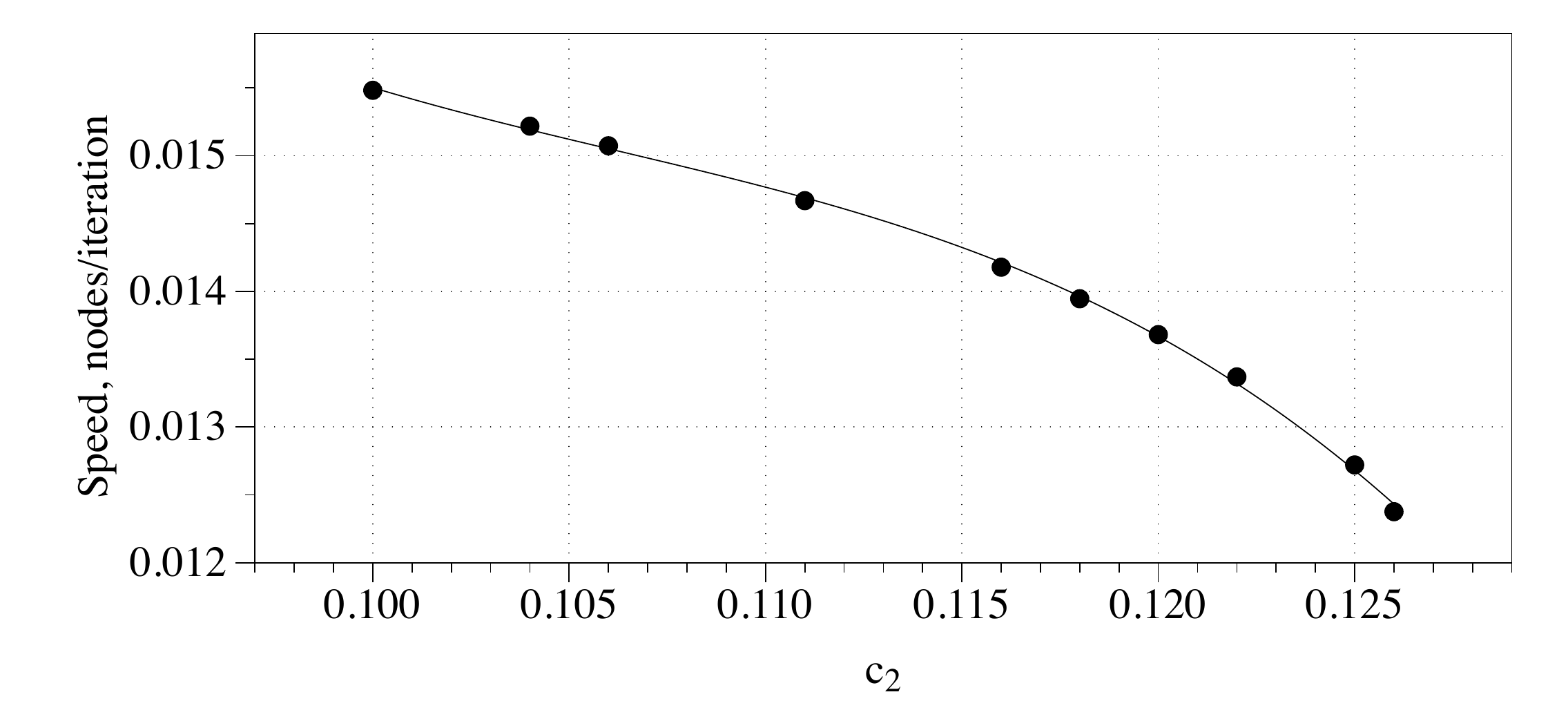}\label{Speed_vs_c2}}
\subfigure[]{\includegraphics[scale=0.43]{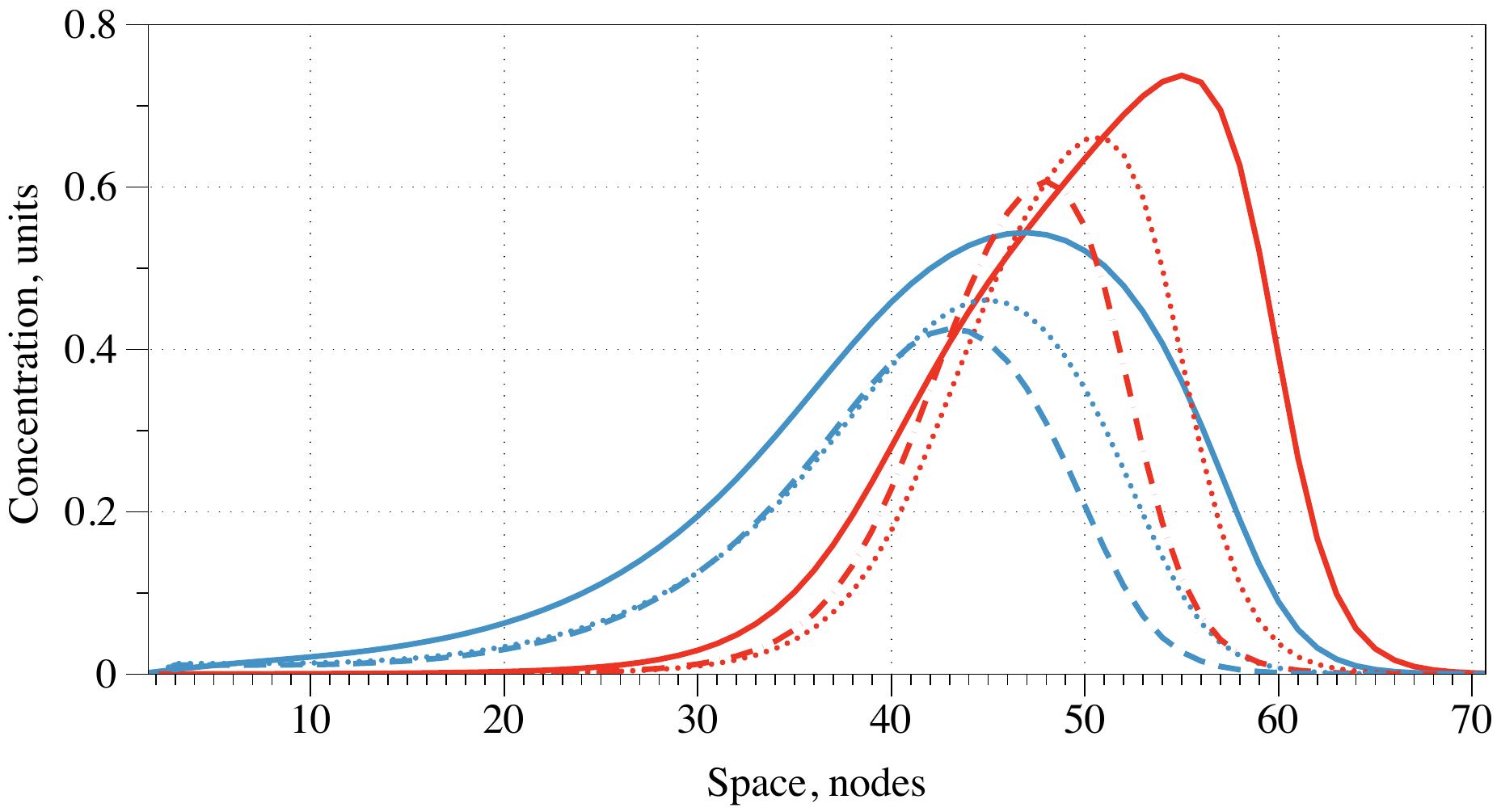}\label{WaveProfile}}
\caption{Propagation of excitation in a free space (a--j) and a channel of 1700$\times$20 nodes (k). Each picture shows not several waves but a configuration of a single wave displayed on the same image every 100\textsuperscript{th} iterations.
(l)~Time lapse of a single wave-front propagating in a channel for $c_2=0.126$.
(m)~Cross cut of the wave-front profile. Concentration of $u$ is shown by red, and $v$ by blue. Solid lines represent the profile of the wave-front for $c_2=0.1$, dotted line for $c_2=0.12$ and dashed line for $c_2=0.126$.
Time lapses (a--j) are at 0.3 scale of the real size, and (k) at 0.27 scale. 
}
\label{fig:expansion}
\end{figure}
\graphicspath{{figs/}}

\paragraph{Propagation of a wave-front in a `free' space.} Let us have a brief look at the wave propagation in a `free' space for selected values of excitability $c_2$ (Fig.~\ref{fig:expansion}). In a fully excitable medium, $c_2=0.1$ we observe a circular wave propagating from a single source of stimulation (Fig.~\ref{expansion0100000}). Since excitability reaches $c_2=0.11$ a stimulation evokes not a circular wave but a wave-fragment. The fragment rapidly expands in Fig.~\ref{expansion0110000}. The time necessary for a wave-fragment to lose its stability and start expanding increases with decrease of excitability (Fig.~\ref{expansion0115000}--\ref{expansion0115090}). The excitation shows  soliton-like behaviour with a wave-fragment staying compact for hundreds of thousands of iterations for $c_2=0.115092$ (Fig.~\ref{nearsoliton0115092}) and $c_2=0.115092$ (Fig.~\ref{soliton0115093}). For $c_2>0.115094$ wave-fragments collapse rapidly.  

\paragraph{Speed of a wave-front propagation in a channel.} 
As seen in Fig.~\ref{collapse0115095}, excitation wave-front collapses after nearly $18.5 \cdot 10^3$ iterations for $c_2=0.115095$, yet when geometrically restricted to a channel of a conducive material, width 20 nodes, the excitation steadily propagates for $c_2$ up to 0.1265 (Fig.~\ref{single_channel_lapse_0126}). Geometrical constraining of an excitation enhances lifetime of wave-fragments at lower values of excitability however does not prevent wave-fragments from slowing down (Fig.~\ref{Speed_vs_c2}). A crude linear dependence of a speed $v$ on excitability $c_2$ is $v(c_2)=0.027041 + (-0.11297) \cdot c_2$ and more accurate cubic one is $v(c_2)=0.21425 + (-5.4905)\cdot c_2 + 51.057\cdot c_2^2 + (-160.27)\cdot c_2^3$. 
\paragraph{Wave profile.} What happens with a wave profile with decrease of excitability? As illustrated in Fig.~\ref{WaveProfile}, the increase of $c_2$ from 0.1 to 0.12 to 0.126 causes decrease in amplitude.  A width of the wave-front does not seem to be affected by the excitability, as seen in Fig.~\ref{WaveProfile}. However, due to a lower amplitude and a narrower shape the wave-fronts are visualised as narrow for higher values of $c_2$.

\paragraph{Excitation propagation into angled branches.}

\begin{figure}[!tbp]
    \centering
    \subfigure[]{\includegraphics[scale=0.55]{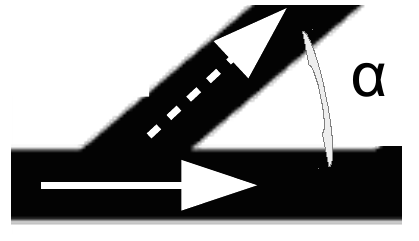}\label{AngleScheme}}\\
    \subfigure[$c_2=0.100$]{\includegraphics[scale=0.27]{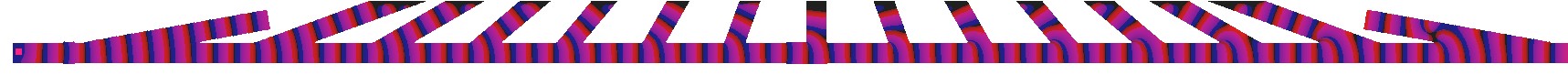}\label{angles0100}}
    \subfigure[$c_2=0.104$]{\includegraphics[scale=0.27]{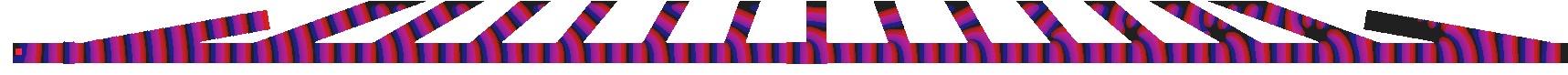}\label{angles0104}}
     \subfigure[$c_2=0.106$]{\includegraphics[scale=0.27]{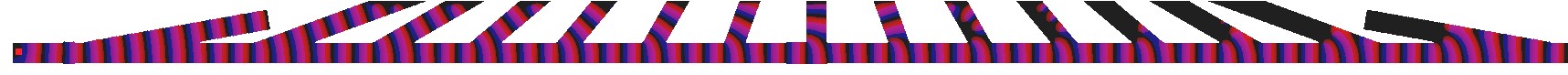}\label{angles0106}}
      \subfigure[$c_2=0.111$]{\includegraphics[scale=0.27]{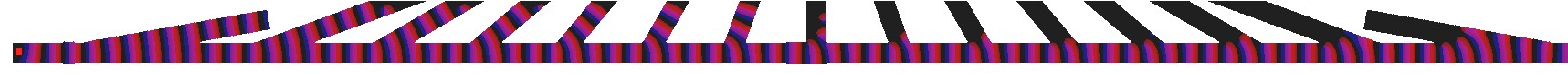}\label{angle0111}\label{angles0111}}
      \subfigure[$c_2=0.118$]{\includegraphics[scale=0.27]{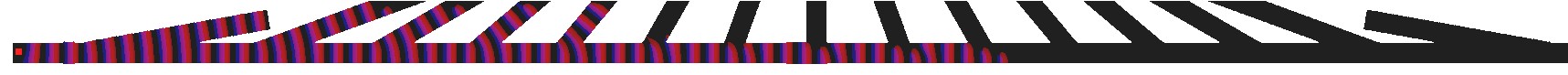}\label{angles0118}}
      \subfigure[$c_2=0.120$]{\includegraphics[scale=0.27]{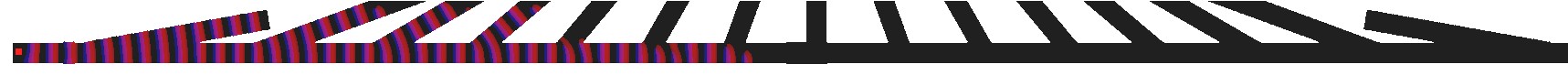}\label{angles0120}}
      \subfigure[$c_2=0.122$]{\includegraphics[scale=0.27]{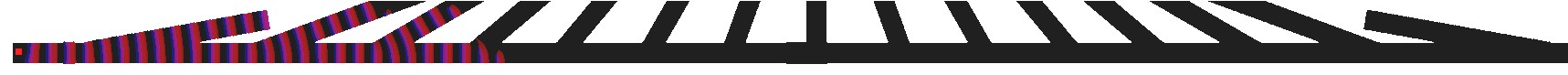}\label{angles0122}}
       \subfigure[$c_2=0.126$]{\includegraphics[width=0.9\textwidth]{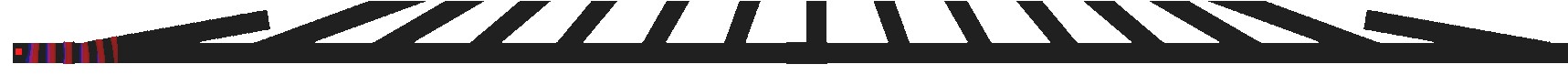}\label{angles0126}}
           \subfigure[]{\includegraphics[scale=0.43]{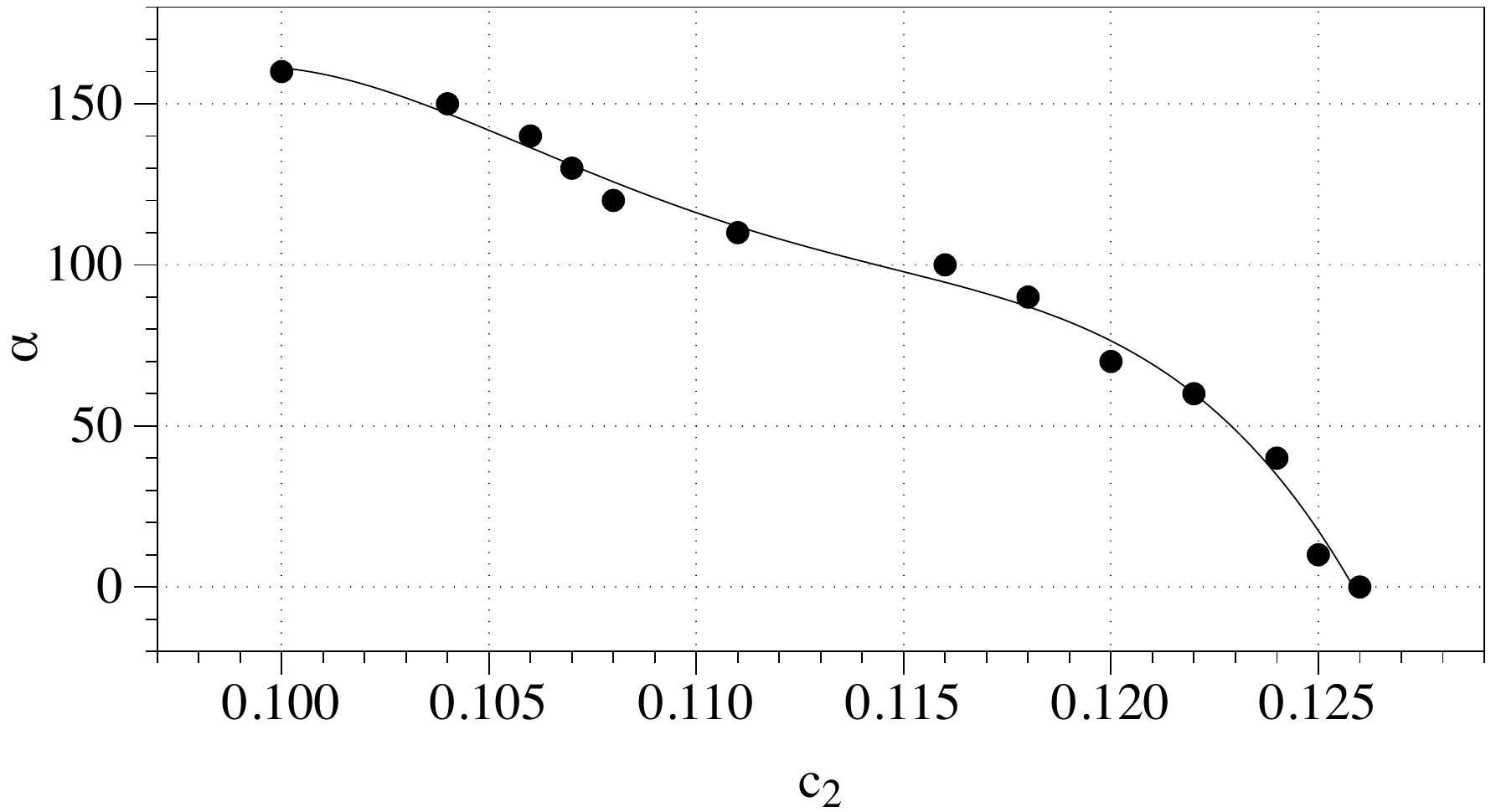}\label{angleVsC2}}
    \subfigure[]{\includegraphics[scale=0.43]{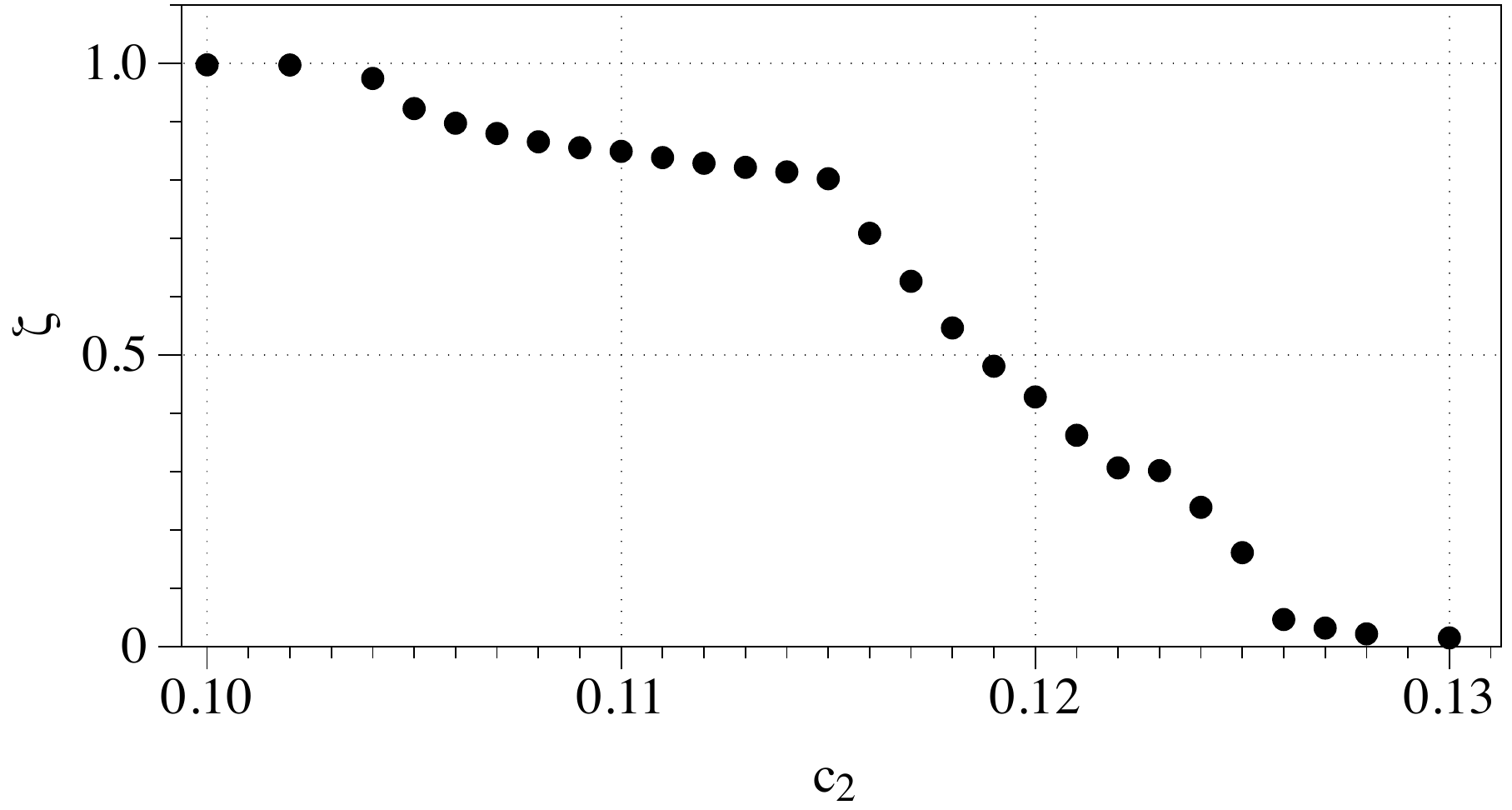}\label{Excitability_vs_Coverage}}
    \caption{Excitability affects a wave-front propagation into angled branches.
    (a)~A representation of the angle $\alpha$. Solid arrow shows direction of propagation of the wave-front in the central channel, dashed arrow shows direction of propagation of the wave-front entering the angled branch. 
    (b--i) Exemplary time lapses of experiments. All channels have width 20 nodes.
    (j)~Angle $\alpha$ of a lateral channel being entered by excitation wave versus $c_2$. Data points are shown by solid discs, polynomial approximations by line. 
    (k)~Coverage $\zeta$ versus non-excitability $c_2$.
    Time lapses (b--i) are at 0.27 scale of their real size. See videos in \cite{adamatzkyFTHZenodo}.
    }
    \label{fig:angles}
\end{figure}

Assume a wave-front propagates along a central channel, 20 nodes wide, from which other channels are branching our at various angles. A width of each lateral channel is 20 nodes. How would excitability of the medium affect an ability of the wave-front to enter a later channel branching at a specified angle? To find the answer we designed a template  with 15 lateral channels branching out the central channel at angles (Fig.~\ref{AngleScheme}) from 20\textsuperscript{o} to 170\textsuperscript{o} with increment 10\textsuperscript{o}. A distance between two adjacent lateral channels is 60 nodes. 

Exemplar time lapses of excitation propagating in the template for various values of $c_2$ is shown in Fig.~\ref{angles0100}--\ref{angles0126}. Excitation enters all lateral channels till $c_2=0.1$ (Fig.~\ref{angles0100}) but fails to enter the channel branching out at 170\textsuperscript{o} when excitability is $c_2=0.104$ (Fig.~\ref{angles0104}). With further increase of $c_2$ less channels branching out at obtuse angles remain unexcited (Fig.~\ref{angles0106}). At excitability level $c_2=0.111$ the excitation fails to enter 90\textsuperscript{o} channel (Fig.~\ref{angles0111}). After further decrease of excitability through $c_2=118$ (Fig.~\ref{angles0118} and $c_2=0.120$ (Fig.~\ref{angles0120}) to  $c_2=0.122$ (Fig.~\ref{angles0120}) excitation fails to propagate along the central channel at $c_2=0.126$.

The largest angle $\alpha$ of a lateral channel entered by an excitation decreases polynomially with increase of $c_2$: $\alpha(c_2)=352532 + (-1.2716\cdot 10^7)\cdot c_2 + 1.7136\cdot 10^8\cdot c_2^2 + (-1.0227 \cdot 10^9)\cdot x^c_2 + 2.2823 \cdot 10^9\cdot c_2^4$ (Fig.~\ref{angleVsC2}). 
In Fig.~\ref{Excitability_vs_Coverage} we give a fine-grained characterisation of the dynamic using the coverage $\zeta$.
The coverage $\zeta$ is a ratio of nodes excited at least once during the simulation to a total number of nodes in the template.  Three critical values of $c_2$ are evident in Fig.~\ref{Excitability_vs_Coverage}:  
$c_2=0.104$ (wave fails to enter 170\textsuperscript{o} lateral channel), 
$c_2=0.116$ (wave fails to enter 80\textsuperscript{o} channel) and 
$c_2=0.126$ (wave fails to propagate in the central channel).

\paragraph{Arrest of an excitation wave-front on its entry into an expansion.}

\begin{figure}[!tbp]
    \centering
   \subfigure[$c_2=0.09$]{\includegraphics[scale=0.6]{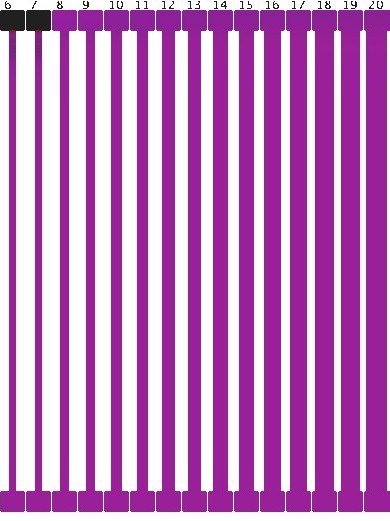}\label{arrest009}}
     \subfigure[$c_2=0.11$]{\includegraphics[scale=0.6]{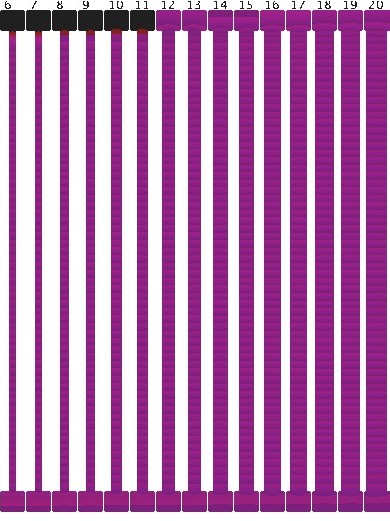}\label{arrest011}}
     \subfigure[$c_2=0.12$]{\includegraphics[scale=0.6]{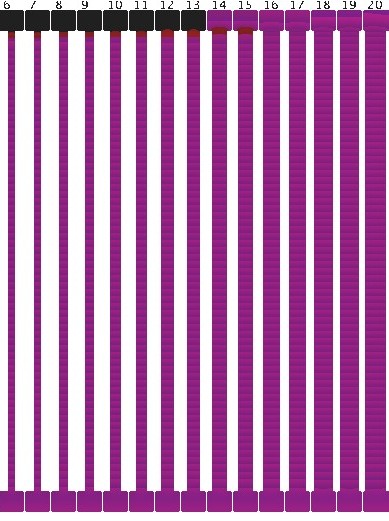}\label{arrest012}}
     \subfigure[]{\includegraphics[scale=0.4]{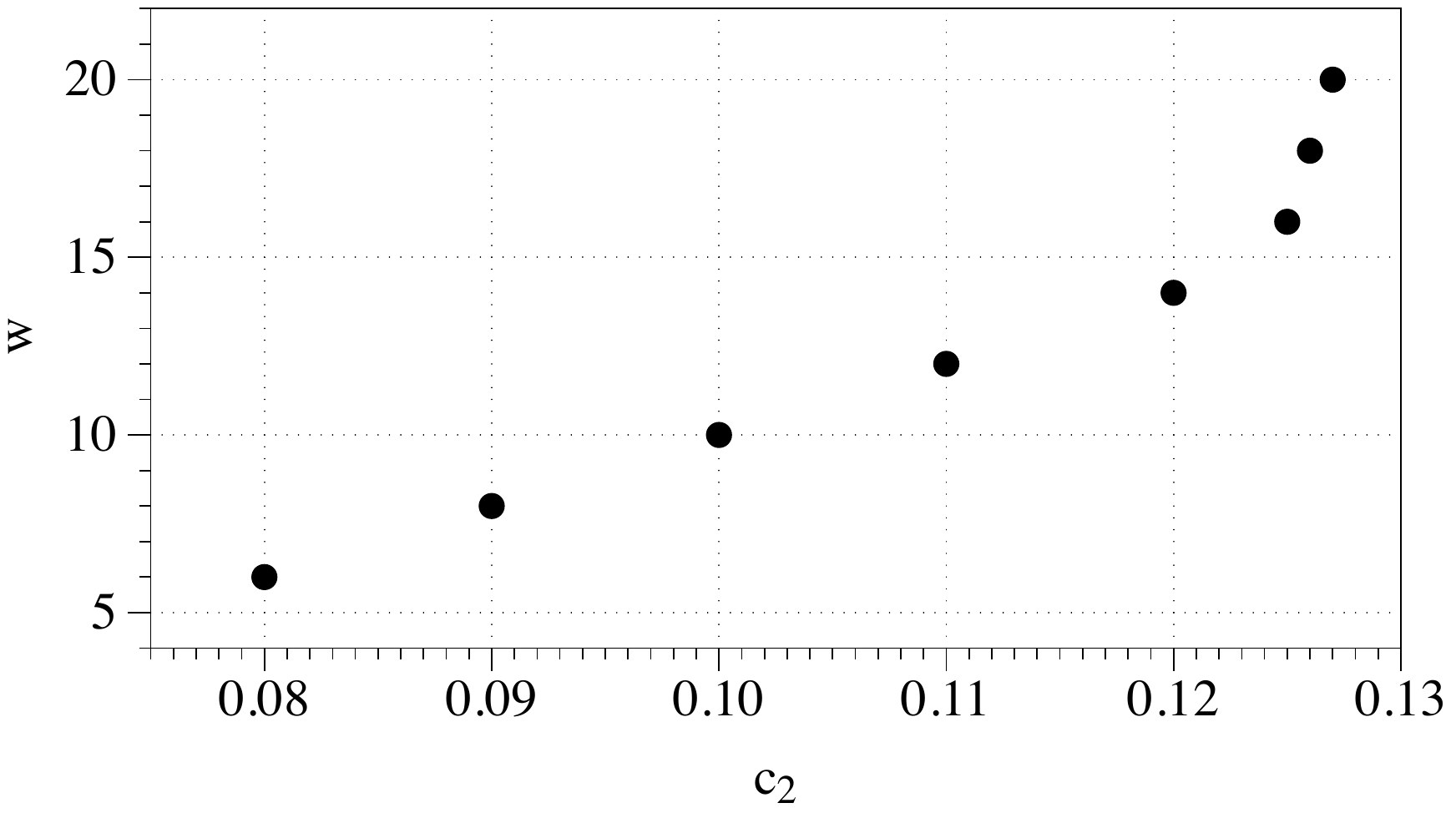}~\label{Width_vs_c2}}
    \caption{Arrest on entering expansion. 
    (abc)~Exemplar time lapse snapshots, scale 0.4 of the real size. Size of expansion in nodes is shown above each template. 
    (d)~$w$ versus $c_2$, where $w$ is the smallest width of a channel from which the excitation propagates into the extension reservoir.}
    \label{fig:arrest}
\end{figure}

To analyse how excitability affects an arrest of a wave-front on entering expansion we prepared 15 templates as follows. Each template is comprised of two rectangles $25 \times 21$ nodes connected by a channel 460 nodes long. Width of the channel was varied from 6 nodes to 20 nodes. We initiated an excitation in the bottom expansion, and allowed the excitation wave to propagate along the narrow channel and into the top expansion. Exemplar time lapses of experiments are shown in Fig.~\ref{arrest009}, \ref{arrest011}, \ref{arrest012}. Let $w$ is the smallest width of a channel from which the excitation propagates into the extension reservoir. The dependence of $w$ on $c_2$ is shown in Fig.~\ref{Width_vs_c2}. The following piece-wise linear approximation is valid: $w(c_2)=2\cdot 10^2 \cdot c_2 - 10$, if  $c_2 \in [0.08, 0.125[$ and $w(c_2)=2\cdot 10^3 \cdot c_2 -234$, if $c_2 \in [0.125, 0.127]$.

\paragraph{Exploration of a random plant graph by excitation wave-fronts.}

\graphicspath{{figs/RandomGraphLapses/}}
\begin{figure}[!tbp]
    \centering
\subfigure[$c_2=0.105$]{\includegraphics[scale=0.14]{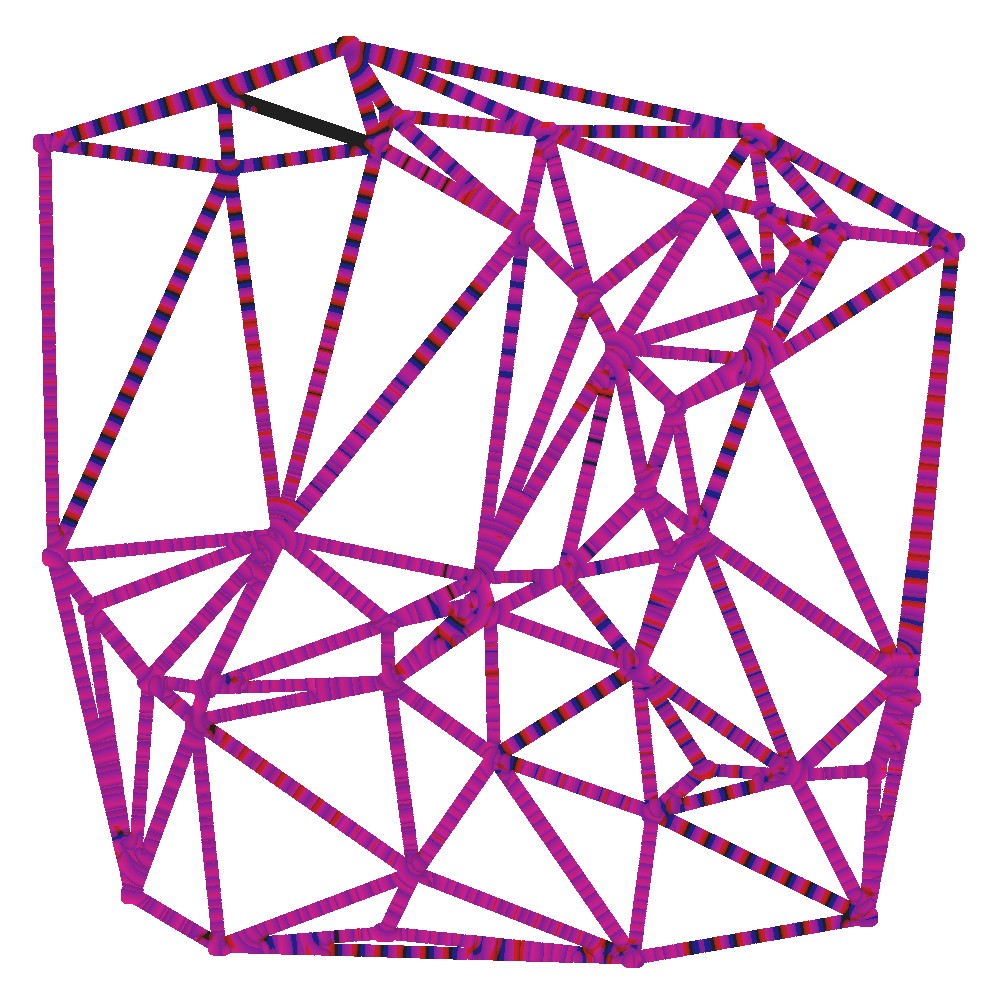}\label{fig:rglapse01050}}
\subfigure[$c_2=0.111$]{\includegraphics[scale=0.14]{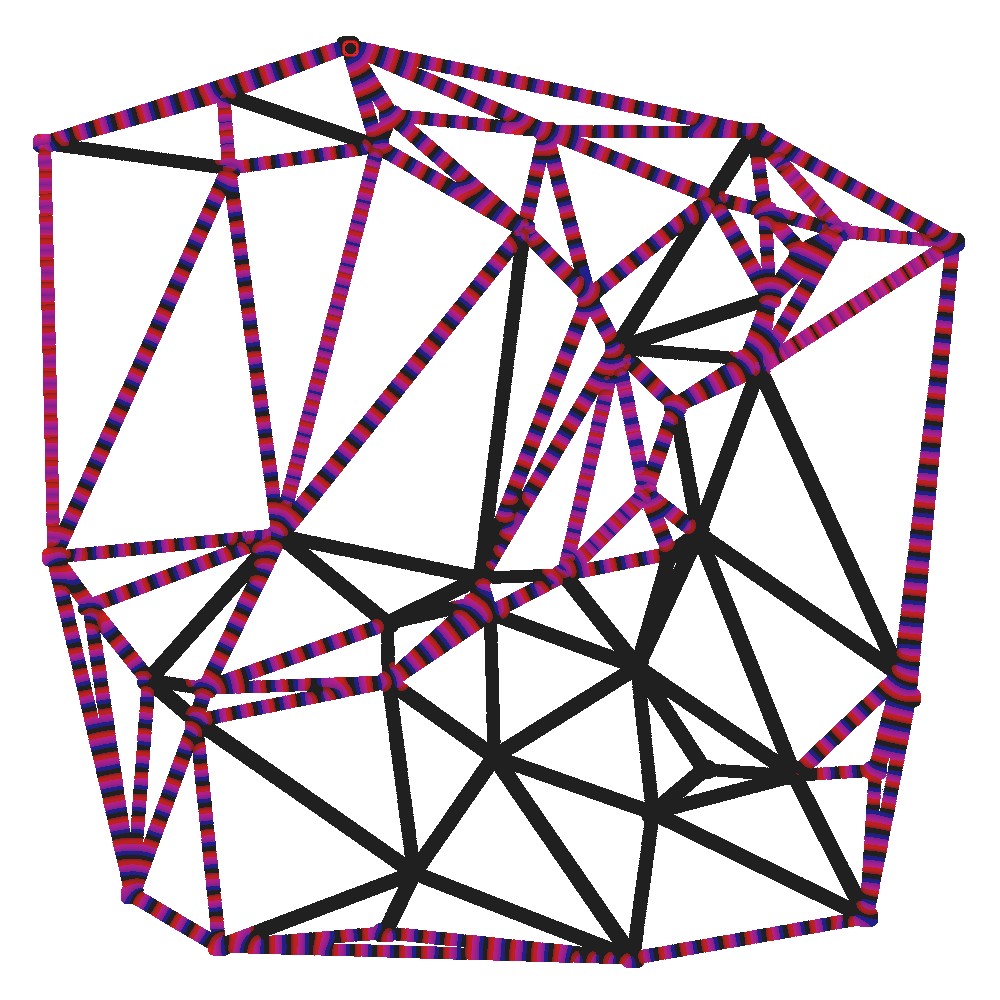}\label{fig:rglapse01110}}
\subfigure[$c_2=0.118$]{\includegraphics[scale=0.14]{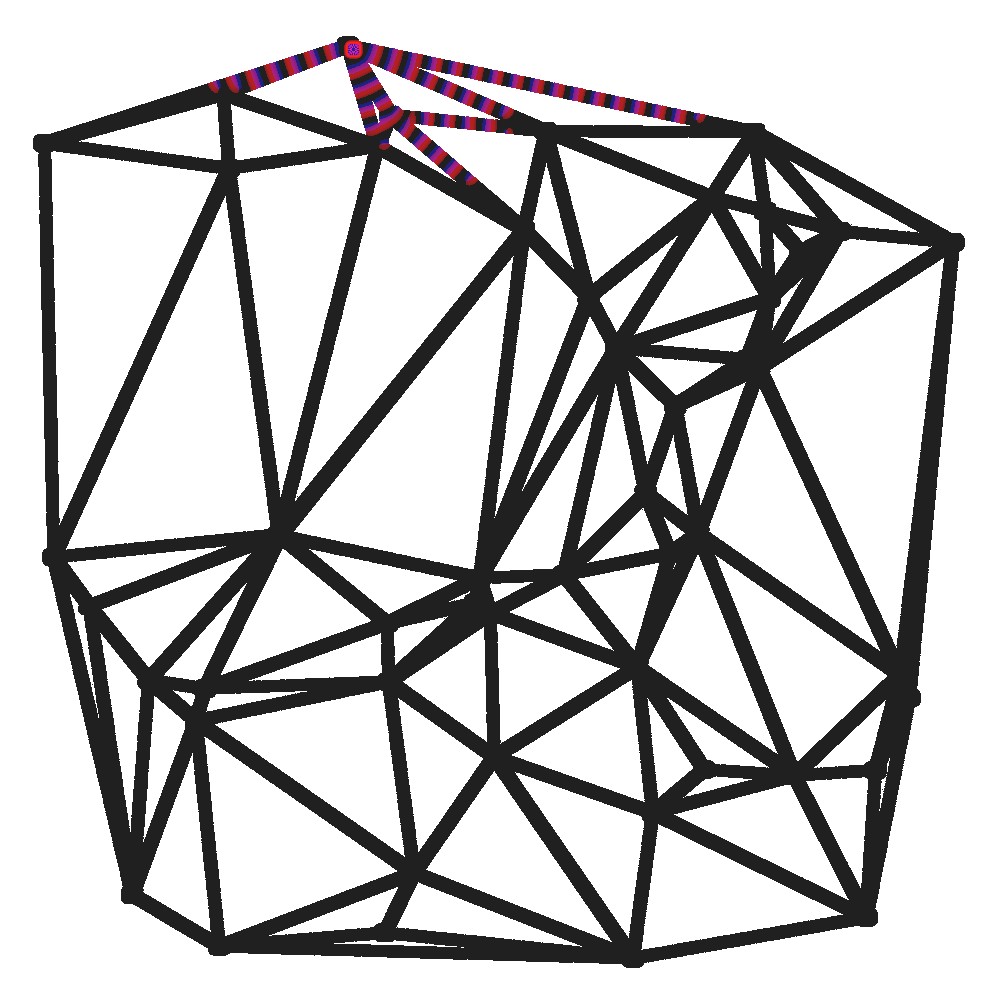}\label{fig:rglapse01180}}
\subfigure[]{\includegraphics[scale=0.43]{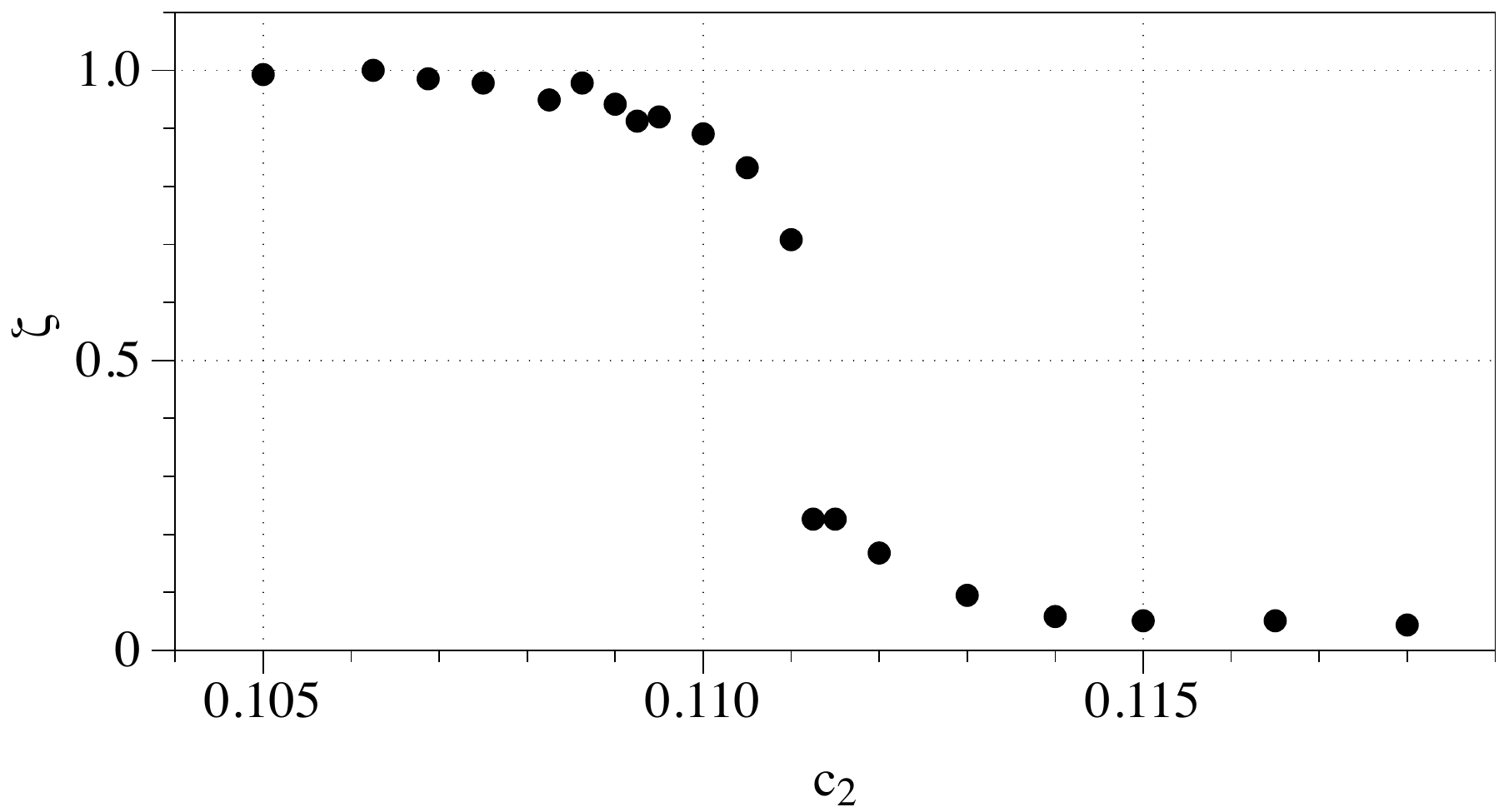}\label{EdgesNonCovered_vs_c2}}
\subfigure[]{\includegraphics[scale=0.43]{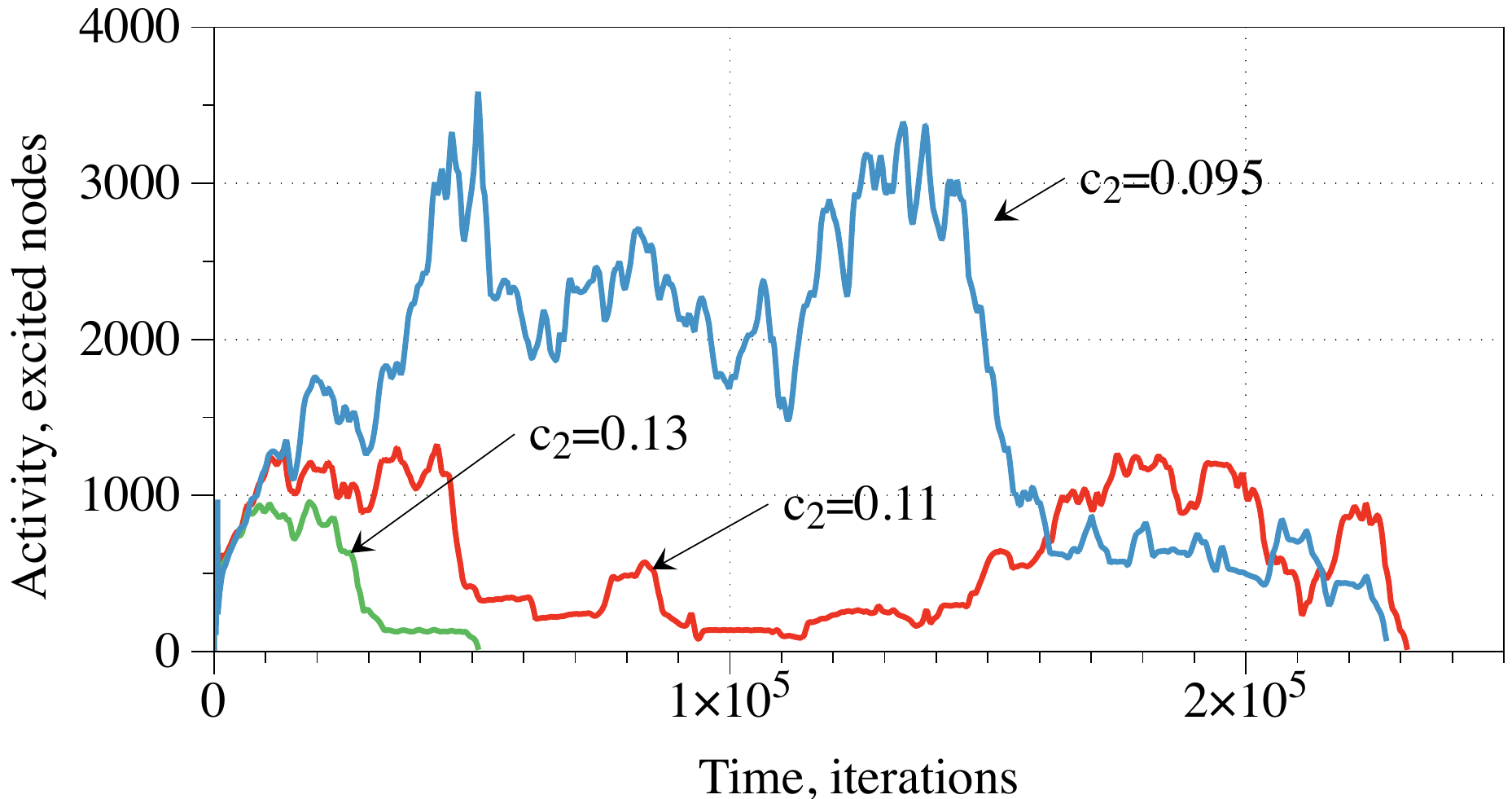}\label{RandomGraphActivity}}
    \caption{Exploration of a random planar graph $\mathcal{G}$ by excitation wave-fronts. (a)--(c)~Time lapse snapshots of the excitations on $\mathcal{G}$. See videos in \cite{adamatzkyFTHZenodo}. 
    (d)~A ratio $\zeta$ of a number of edges traversed by excitation wave-fronts to a total number of edges in $\mathcal{G}$  versus $c_2$. }
    \label{fig:randomgraph}
\end{figure}
\graphicspath{{figs/}}

 A random planar graph $\mathcal{G}$ is generated using  Grapherator R package~\cite{grapherator} as a random set of 50 planar nodes (samples points from a bi-variate  uniform  distribution) connected by 137 edges of a Delaunay triangulation.  For each of the tested values of $c_2$ we started with a graph in a resting state and stimulated the north-most node of the graph.  Exemplar time lapses of excitation on $\mathcal{G}$ are shown in Fig.~\ref{fig:randomgraph}. From the plot of ratio $\zeta$ of edges traversed by excitation wave-fronts versus $c_2$ we see that $\zeta$ decreases with increase of $c_2$. At $c_2=0.111$ we witness a sharp decrease in graph coverage (Fig.~\ref{EdgesNonCovered_vs_c2}). Note that at the same value of $c_2$ excitation wave-fronts are no longer entering branches with angle over 90\textsuperscript{o} (Fig.~\ref{angle0111}) and chances of arrest on entering an expansion start to increase exponentially with growth of $c_2$ (Fig.~\ref{Width_vs_c2}). The coverage $\zeta(c_2)$ can be approximated by two cubic functions: 
$4125.3 -115892\cdot c_2 + 1.0855\cdot 10^6 \cdot c_2^2 -3.3891 \cdot 10^6 \cdot c_2^3$ for $c_2 \in [0.105, 0.111]$ and $2442.2 -62991) \cdot c_2 + 541558\cdot c_2^2 -1.5519\cdot 10^6 \cdot c_2^3$ for  $c_2 \in ]0.111, 0.118]$. As evidenced by videos of experiments~\cite{adamatzkyFTHZenodo} and exemplar plots of activity in Fig.~\ref{RandomGraphActivity} the traversing of the $\mathcal{G}$ by excitation wave-fronts is non-linear. For $c_2$ below 0.09 we see multiplications of wave-fronts, with some fronts travelling along cyclic routes on the graph. Few outburst of activity, due to multiplication of the fronts for $c_2=0.095$ is shown in Fig.~\ref{RandomGraphActivity}. For $c_2=0.11$ we witness prolonged periods of low activity with few outbursts of higher aperiodic activity. This is due to a small number of wave-fronts traversing the graph. Some collisions between the wave-fronts lead to annihilation of the fronts, others to formation of additional wave-fronts. 

\section{Discussion}

\begin{figure}[!tbp]
    \centering
    \includegraphics[scale=0.47]{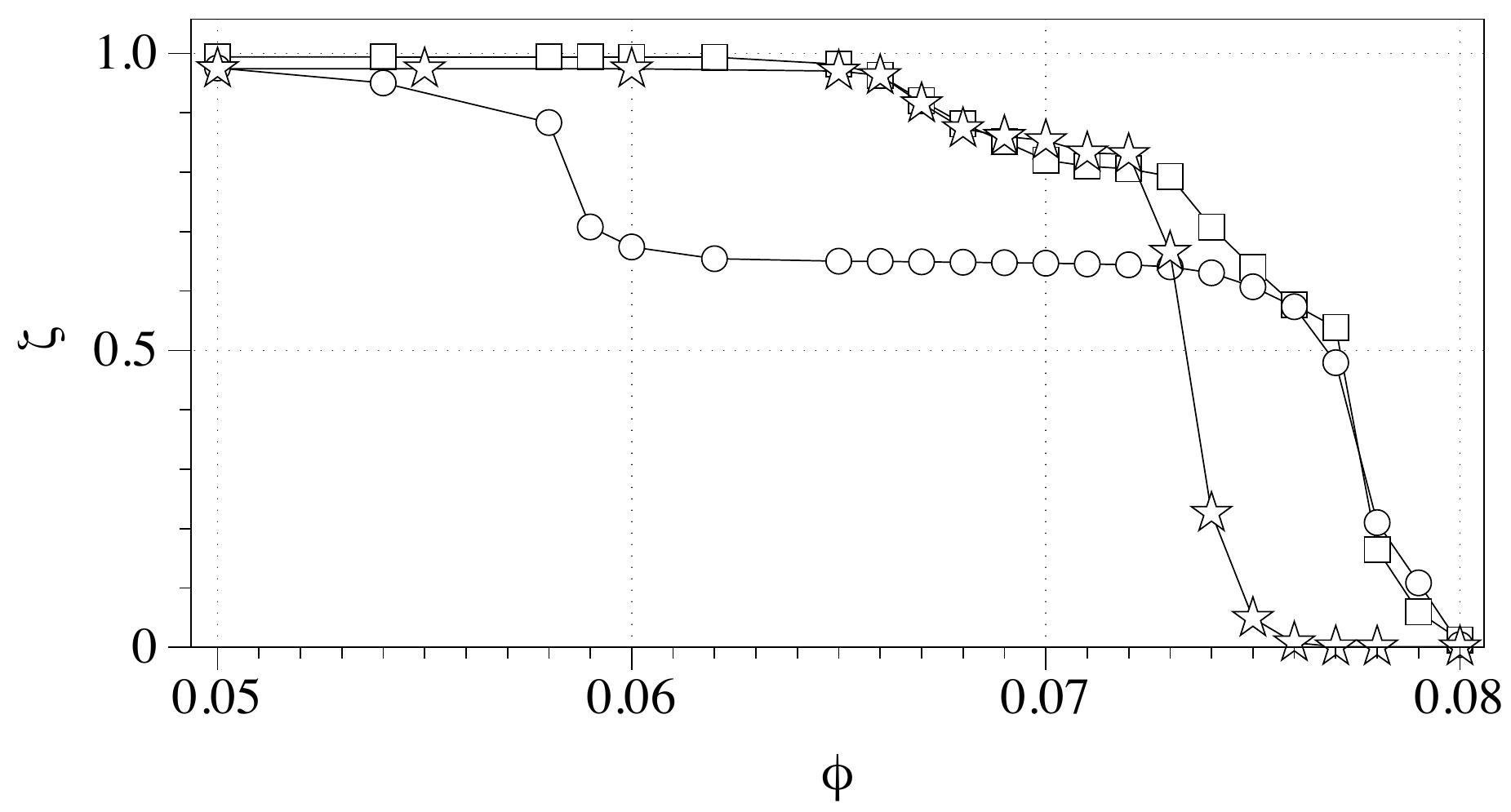}
    \caption{Coverage $\zeta$ of selected fragments of street networks in Tehran (circle), Barcelona (star), London (square) versus excitability $\phi$ of the two-variable Oregonator model of Belousov-Zhabotinsky reaction. See detials in \cite{adamatzkyBarcelona}. }
    \label{fig:cities}
\end{figure}

We found how excitability affects a speed of excitation wave-fronts, an ability of the fronts to expand into angled branches, and chances of the fronts to be arrested on entering an expansion. While studying exploration of a random planar graph with excitations we found that the coverage as a function of excitability is a sigmoid function, and that for certain range of excitability values we can observe repeated outburst of excitations due to the wave-fronts travelling along cyclic paths and producing multiple wave-fragments while interacting with each other. The function is comparable with coverage functions obtained in our numerical experiments with two-variable Oregonator model of Belousov-Zhabotinsky (BZ) reaction, where excitability is controlled by parameter $\phi$, and city streets networks are explored by oxidation wave fronts: Barcelona~\cite{adamatzkyBarcelona}, London~\cite{adamatzky2018street} and Tehran~\cite{adamatzky2018exploring} (Fig.~\ref{fig:cities}). Despite different equations used (Oregonator for street networks and FHN for a random planar graph) similarities are striking. In both cases we have a sigmoid function. The function is rather perfect in case of a random graph (Fig.~\ref{EdgesNonCovered_vs_c2})  and the function is less pronounced, or even distorted, in the case of street networks (Fig.~\ref{fig:cities}). In the case of a random graph the sharp transition starts roughly at the middle, $c_2=0.1115$ of the interval $c_2=0.105$, full coverage, and $c_2=0.118$, no coverage. Also, noticeably, the coverage function for Tehran shows two sharp drops in coverage, first just near $\phi=0.06$ and second near $\phi=0.075$.  Further studies are required to establish deep relationships between Oregonator BZ and FHN models' behaviour on graphs and their potential for analysing the geometric networks.

\bibliographystyle{plain}
\bibliography{bibliography}

\end{document}